\documentclass{aa}

\usepackage{txfonts}
\usepackage{graphicx}
\usepackage{natbib}

\newcommand{\arXiv}{arXiv}

\newcommand{\integral}{INTEGRAL}
\newcommand{\swift}{\emph{Swift}}
\newcommand{\chandra}{\emph{Chandra}}
\newcommand{\spitzer}{\emph{Spitzer}}

\newcommand{\glimpse}{GLIMPSE}

\newcommand{\arcdeg}{$^{\circ}$}

\newcommand{\xtesixteen}{XTE\,J1637$-$498}
\newcommand{\igrseventeen}{IGR\,J17379$-$3747}

\newcommand{\igrseventeenfive}{IGR\,J17585$-$3057}
\newcommand{\gxnine}{GX\,9+1}
\newcommand{\sgrxthree}{Sgr\,X$-$3} 

\newcommand{\onea}{1A\,1742$-$294}
\newcommand{\slxoo}{SLX\,1744$-$300}
\newcommand{\igrseventeenofive}{IGR\,J17505$-$2644}

\newcommand{\lmxb}{LMXB}
\newcommand{\lmxbs}{LMXBs}

\begin{document}


\title{Discovery and identification of  infrared counterpart candidates of four Galactic centre low mass X-ray binaries\thanks{Based on observations collected at the European Organisation for Astronomical Research in the Southern Hemisphere, Chile under ESO programs 081.D-0401 \& 084.D-0535 (P.I. Chaty)}}


\titlerunning{IR counterpart candidates of four Galactic centre LMXBs}

  \author{
    P.A.~Curran\inst{1}
    \and S.~Chaty\inst{1}
    \and J.A.~Zurita Heras\inst{2}
  }


  \offprints{P.A.~Curran (peter.curran@cea.fr)}
  
  \institute{
    Laboratoire AIM, CEA/IRFU-Universit\'e  Paris Diderot-CNRS/INSU, CEA DSM/IRFU/SAp, Centre de Saclay, F-91191  Gif-sur-Yvette, France 
    \and Fran\c{c}ois Arago Centre, APC, Universit\'e Paris Diderot, CNRS/IN2P3, CEA/DSM, Observatoire de Paris, 13 rue Watt, 75205 Paris Cedex 13, France 
  }



\abstract
{The near infrared (nIR)/optical counterparts of low mass X-ray binaries (\lmxbs) are often observationally dim and reside in high source density fields which make their identification problematic; however, without such a counterpart identification we are unable to investigate many of the properties of \lmxb\ systems.}
%
{Here, in the context of a larger identification campaign, we examine the fields of four \lmxb\ systems near the Galactic centre, in a bid to identify nIR/optical counterparts to the previously detected X-ray point sources.}
%
{We obtain nIR/optical images of the fields with the ESO -- New Technology Telescope and apply standard photometric and astrometric calibrations; these data are supplemented by \spitzer -\glimpse\ catalog data. } 
%
{On the basis of positional coincidence with the arcsecond accurate X-ray positions,  we identify unambiguous counterpart candidates for \object{\xtesixteen}, \object{\igrseventeen}, \object{\igrseventeenfive}\ and \object{\gxnine}. }
%
{We propose tentative nIR counterparts of four \lmxbs\  which require further investigation to confirm their associations to the X-ray sources. }

\keywords{ 
  X-rays: binaries --  infrared: stars 
  -- X-rays: individuals: \xtesixteen, \igrseventeen, \igrseventeenfive, \gxnine 
}

\maketitle

\section{Introduction}\label{section:intro}

Given the main sequence or red giant nature of the companion stars of low mass X-ray binaries (\lmxbs), they are intrinsically dim in the  optical and near-infrared (nIR), and hence difficult to detect (see e.g., \citealt {Charles2003:astro.ph.802} for a review of the optical and nIR properties of \lmxbs). The intrinsic dimness of all these sources is further compounded by the high level of optical extinction in the direction of the Galactic plane \citep{schlegel1998:ApJ500}, where many \lmxbs\ are found \citep{Grimm2002:A&A391}; hence observing at nIR wavelengths, where this is less pronounced, may increase the chance of a detection. 
Due to their positions in the Galactic plane or centre, where the optical/nIR (OIR) source density is extremely high, accurate X-ray or radio positions are necessary to confidently identify the OIR counterparts of \lmxbs\ and reduce the probability of chance superpositions. The required (sub)arcsecond positions are now more readily available due to missions such as \swift, XMM-{\it Newton} and \chandra, and when combined with the current generation of optical telescopes and instruments, increase the likelihood of counterpart discovery.

If the \lmxb\ is a persistent source or a transient source in outburst \citep{vanParadijs1994:A&A.290} emission from the active accretion disk may be brighter than the companion star and more likely to be detected at OIR wavelengths; at nIR wavelengths it is even possible that a jet component will make a significant contribution to the emission (e.g., \citealt{Russell2010:arXiv1001}).
Broadband SED fitting of observed OIR counterparts can allow the various emission components (companion star and accretion disk/jet if present) to be disentangled from each other and, along with spectral lines, allows the spectral type -- and hence mass -- of the companion to be constrained. When combined with radial velocity information from X-ray or optical light curves this allows constraints to be placed on the mass of the compact object with which to confirm the nature of the compact object as derived from X-ray spectral or timing properties.  Furthermore the relationship between the OIR and X-ray flux can be used as a diagnostic with which to derive the compact object or companion star nature, and to identify the state of the system at the time of observations (e.g., \citealt{russell2006:MNRAS371}).

\begin{table*}	
  \centering	
  \caption{X-ray and nIR positions for the \lmxbs\ in our study} 	
  \label{table:positions} 	
  \begin{tabular}{l l l l l l l l} 
    \hline\hline
     & X-ray &  &  & & nIR &  &  \\ 
    Source & RA & Declination & Err (\arcsec) & & RA & Declination & Err (\arcsec) \\ 
    \hline 
    \xtesixteen$^1$ & 16:37:02.67  & $-$49:51:40.6 & 1.8  & & 16:37:02.66  & $-$49:51:40.0 & 0.16 \\
    \igrseventeen$^2$ & 17:37:58.81 &  $-$37:46:19.6 & 1.4 & &  17:37:58.77 &  $-$37:46:19.9 & 0.32 \\
    \igrseventeenfive$^3$ & 17:58:29.85 &  $-$30:57:01.6 & 0.6 & &  17:58:29.83 &  $-$30:57:01.6 & 0.16 \\
    \gxnine$^3$ & 18:01:32.15 &  $-$20:31:46.1 & 0.6 & &  18:01:32.15 &  $-$20:31:46.2 & 0.16 \\
    \hline 
  \end{tabular}
  \begin{list}{}{}
  \item[]  Reference to X-ray positions: $^1$ \cite{Starling2008ATel.1704}, $^2$ \cite{Krimm2008:ATel.1714}, $^3$ M. Revnivtsev (private communication)  
  \end{list}
\end{table*}

Here we examine the fields of four \lmxb\ systems near the Galactic centre, in a bid to identify nIR/optical counterparts to the previously detected X-ray point sources (Table\,\ref{table:positions}). For \xtesixteen, \igrseventeen, \igrseventeenfive\ and \gxnine\ we identify unambiguous counterpart candidates on the basis of positional coincidence with the arcsecond accurate X-ray positions. 
We have also observed a further three \lmxb\ systems (\object{\onea}, \object{\slxoo}\ and  \object{\igrseventeenofive})$^*$ as part of our campaign to identify the nIR counterparts of high energy sources near the Galactic centre. Though we do not detail these observations, they confirm the results of \cite{Zolotukhin:2011MNRAS.411} and have been used to verify our reduction methods. 
In section \ref{section:observations} we introduce our observations and reduction method while in section \ref{section:results}, after briefly introducing each source,  we detail the results of those observations.  We summarise our findings in section \ref{section:conclusions}.  Throughout, positions (J2000) are given with 90\% confidence while all others values, including magnitudes, are given with $1\sigma$ confidence.


\section{Observations and data analysis}\label{section:observations}

Optical  ($U,B,V,R,i$) and nIR ($J,H,K_s$) data were obtained with the ESO Faint Object Spectrograph and Camera (v.2; EFOSC2) and the Son of ISAAC (SofI) infrared spectrograph and imaging camera on the $3.58$m ESO -- New Technology Telescope (NTT) over two observing runs in September/October 2008 and March 2010 (Table\,\ref{table:dates}). 
The NTT data used a dithered pattern of 3 and 9 exposures per final image in the optical and nIR respectively. Data were reduced using the {\small IRAF} package wherein crosstalk correction, flatfielding, sky subtraction, bias-subtraction and frame addition were carried out as necessary. 
The $3.9\arcmin \times 3.9\arcmin$ images were astrometrically  calibrated against 2MASS \citep{Skrutskie2006:AJ.131} or USNO-B1.0 \citep{Monet2003:AJ125} within the GAIA package and given positional errors include a 0.16\arcsec\ 2MASS systematic uncertainty (Table\,\ref{table:positions}).

PSF photometry was carried out on the final images using the {\small DAOPHOT} package \citep{stetson1987:PASP99} within {\small IRAF}.  The magnitude of the source of interest in each field was calculated relative to a number of comparison stars in the field, including the scatter as a measure of error. The comparison stars were calibrated against \cite{Persson1998:AJ116} or \cite{Landolt1992:AJ.104} photometric standards (though in the case of  \igrseventeen, no suitable optical standards were observed so magnitudes were calibrated against USNO-B1.0 sources). 
If a field was observed on more than one night we tested for variability, using only the error associated with the relative magnitude. As no variability was found, the tabulated magnitudes are calculated from the weighted average relative magnitudes and, since the exposures being compared were equal, the quoted exposures are of a single image only.
The equation, $i-I = (0.247 \pm 0.003) (R-I)$ \citep{Jordi2006A&A.460}, was used to transform the cataloged $I$ magnitudes of the standard stars into $i$ magnitudes with which to calibrate the images. It was again used to transform the observed $i$ band magnitude of the object into appropriate $I$ magnitudes.  Upper limits are approximated from the dimmest observable object in the region of interest.

\begin{table}	
  \centering	
  \caption{Observation log} 	
  \label{table:dates} 	
  \begin{tabular}{l l l} 
    \hline\hline
    Source &  Night of observation & Filters \\ 
    \hline 
    \xtesixteen       & September 9  2008     & optical/nIR\\
                      & September 20, 21 2008 & optical, nIR\\
    \igrseventeen     & September 29 2008     & optical/nIR\\
                      & October 2, 3 2008     & optical, nIR\\
    \igrseventeenfive & March 29 2010         & $K_S$\\
    \gxnine           & March 28 2010         & $K_S$\\
    \hline 
  \end{tabular}
  \begin{list}{}{}
  \item[] Note that the second epoch of observations, if done, were carried out over 2 nights: optical on the first, nIR on the second.
  \end{list}
\end{table}


To supplement our own observations, we utilise data from the \emph{Spitzer Space Telescope}'s  \citep{Werner2004:ApJS154} Galactic Legacy Infrared Mid-Plane Survey Extraordinaire (\glimpse; \citealt{Benjamin2003:PASP115}), when available. \glimpse\  was carried out by the IRAC instrument \citep{Fazio2004:ApJS154} aboard \spitzer, spans 65\arcdeg\ either side of the Galactic center up to $\pm$2-4\arcdeg\ in latitude and includes mosaic images and catalog entries at 3.6, 4.5, 5.8 and 8.0 $\mu$m.


The observed infrared magnitudes  (Table\,\ref{table:mags}) were first converted to flux densities, $F_{\nu}$, at frequency $\nu$,  then to  flux per filter, $F_{filter}$ in units of photons\,cm$^{-2}$\,s$^{-1}$. 
This is done via  $F_{filter} =  1509.18896  F_{\nu}$ $( \Delta\lambda/\lambda )$ where $\lambda$ and $\Delta\lambda$ are the effective wavelength and full width at half maximum of the filter in question. 
{\tt XSPEC} compatible files, for Spectral Energy Distribution (SED) fitting, are then produced from the flux per filter value  using the {\tt FTOOL}, {\tt flx2xsp}.

\begin{table}	
  \centering	
  \caption{Apparent magnitudes of the \lmxb\ counterparts} 	
  \label{table:mags} 	
  \begin{tabular}{l r l} 
    \hline\hline
    Source & & \\ 
    Filter & Exp (s) & Magnitude \\ 
    \hline 
    \xtesixteen$\dagger$ & & \\
    $U$	  & 180 & 21.51 $\pm$ 0.14  \\
    $B$	  & 180 & 20.77 $\pm$ 0.05  \\
    $V$	  & 180 & 18.90 $\pm$ 0.05  \\
    $R$	  & 180 & 17.66 $\pm$ 0.05  \\
    $I$	  & 180 & 16.24 $\pm$ 0.16  \\
    $J$	  & 90 & 14.80 $\pm$ 0.09  \\
    $H$	  & 90 & 14.13 $\pm$ 0.10  \\
    $K_S$ & 90 & 13.55 $\pm$ 0.07  \\
    3.6\,$\mu$m  & ... & 12.87 $\pm$	0.09	\\
    4.5\,$\mu$m  & ... & 12.51 $\pm$	0.10	\\
    5.8\,$\mu$m  & ... & 12.41 $\pm$	0.17	\\
    8.0\,$\mu$m  & ... & 12.28 $\pm$	0.24   \\
    \hline 
    \igrseventeen$\dagger$ & &  \\
    $R$	 & 180 & 21.5 $\pm$ 0.3  \\
    $I$	 & 180 & 20.0 $\pm$ 0.4  \\
    $J$	 & 90 & 18.3 $\pm$ 0.1  \\
    $H$	 & 90 & $\gtrsim 18.0$  \\
    $K_S$& 90 & $\gtrsim 18.0$ \\
    \hline 
    \igrseventeenfive & & \\
    $K_S$ & 54 & 14.18  $\pm$ 0.05 \\
    3.6\,$\mu$m  & ... & 11.38 $\pm$	0.05	\\
    4.5\,$\mu$m  & ... & 10.71 $\pm$	0.05	\\
    5.8\,$\mu$m  & ... & 10.07 $\pm$	0.05	\\
    8.0\,$\mu$m  & ... & 9.30  $\pm$	0.03   \\
    \hline 
    \gxnine & & \\
    $K_S$ & 90 & 15.35 $\pm$ 0.04 \\
    \hline 
  \end{tabular} 
  \begin{list}{}{}
  \item[$\dagger$] For observations with 2 epochs, magnitudes were consistent so we present the weighted average and, since the exposures being compared were equal, the exposure of a single image. 
  \end{list}
\end{table}


\section{Results}\label{section:results}

 \begin{figure*} 
  \centering 
  \resizebox{\hsize}{!}{\includegraphics[angle=-0]{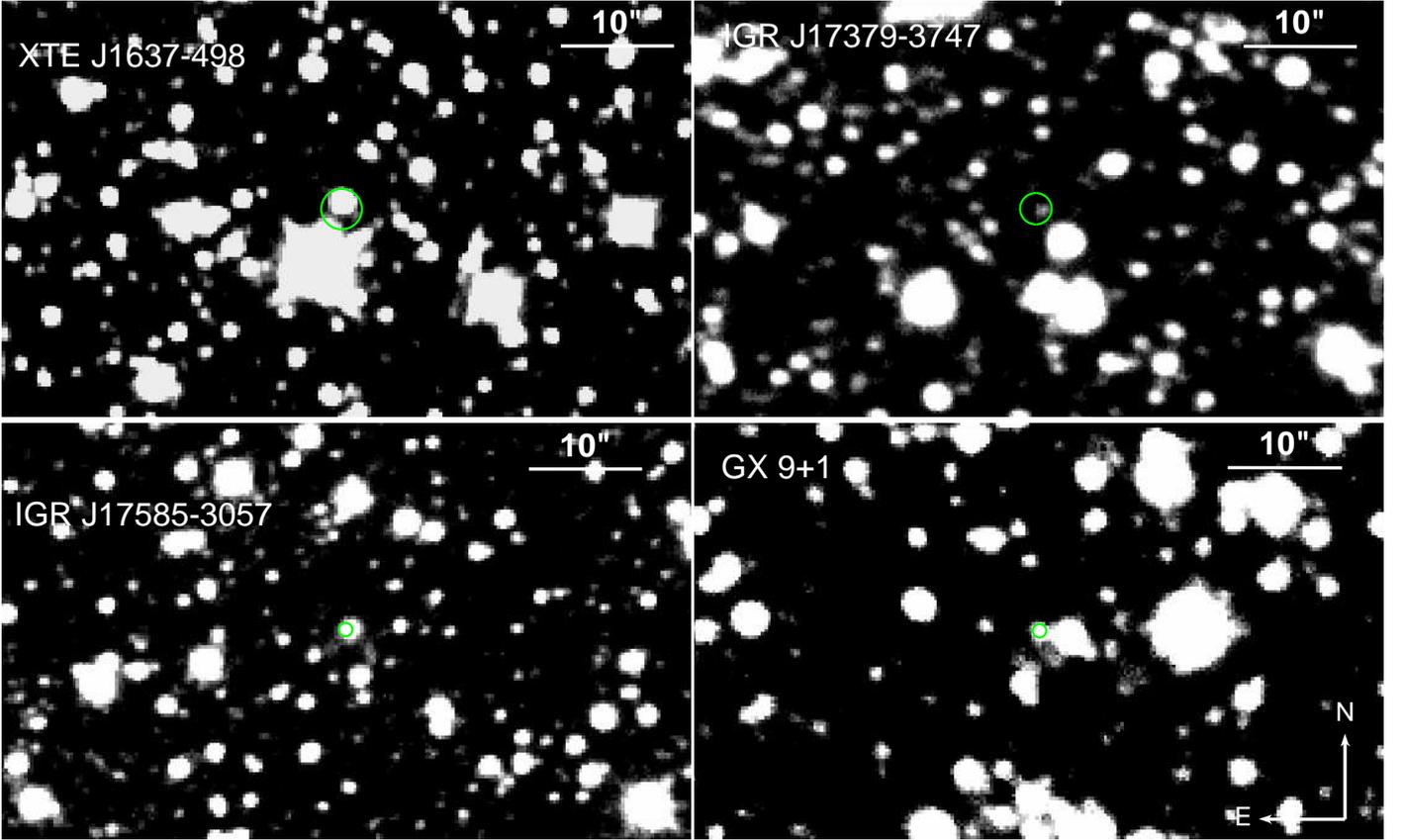}}
  \caption{NTT images ($60\arcsec \times 35\arcsec$) of the fields of \xtesixteen, \igrseventeen, \igrseventeenfive\  and \gxnine\ with the X-ray 90\% position marked. \igrseventeen\ is an EFOSC2 $R$  band image, while all others are SofI $K_S$  band images; all have a common scaling and orientation. }
  \label{fig.4fields} 
\end{figure*}

\subsection{\xtesixteen}

The transient X-ray source, \xtesixteen, was first detected by the \emph{Rossi X-ray Timing Explorer} (RXTE) at a 2-10\,keV flux of 2-4 mCrab; an outburst which lasted approximately from August 25 to September 5 2008 \citep{Markwardt2008ATel.1699}. 
Follow up observations by the \swift\ X-ray Telescope (XRT) allowed the position to be refined (Table\,\ref{table:positions}; \citealt{Starling2008ATel.1704}). Based on a power law fit (photon index  $1.5 \pm 0.4$) of the  XRT spectrum, \cite{Wijnands2008ATel.1700} suggested that the source was an \lmxb, though they cautioned that another system type (i.e., high mass X-ray binary) could not be excluded. \citeauthor{Starling2008ATel.1704} also  noted that  an optical source, consistent with 2MASS object J16370267$-$4951401, was detected in the $v$ band by the \swift\ Ultraviolet \& Optical Telescope (UVOT), within the XRT positional error. No further analysis of this source was published so its true classification remains unknown.

Within the X-ray positional error, we find that an optical and nIR source is detected in all filters (Figure\,\ref{fig.4fields}; best seeing was $\sim 0.95$\arcsec\ in nIR filters and 1.0--1.1\arcsec\ in optical filters).  From examination of our second epoch nIR  images we derive a position of RA, Dec = 16:37:02.66 $-$49:51:40.0 ($\pm 0.16$\arcsec), consistent with 2MASS object J16370267$-$4951401, noted by \cite{Starling2008ATel.1704}.
We find no variability ($\lesssim0.05$ magnitudes) of the counterpart between the two epochs so present only the average values (Table\,\ref{table:mags}). 
We note that the observed magnitudes in $J$, $H$ and $K_s$ are consistent with the cataloged 2MASS values; this is not surprising, given that the observations were obtained after the X-ray outburst had faded. Furthermore we note that this source is also tabulated in the \glimpse\ catalog (G335.4256-01.7828).

\begin{figure} 
  \centering 
  \resizebox{\hsize}{!}{\includegraphics[angle=-90]{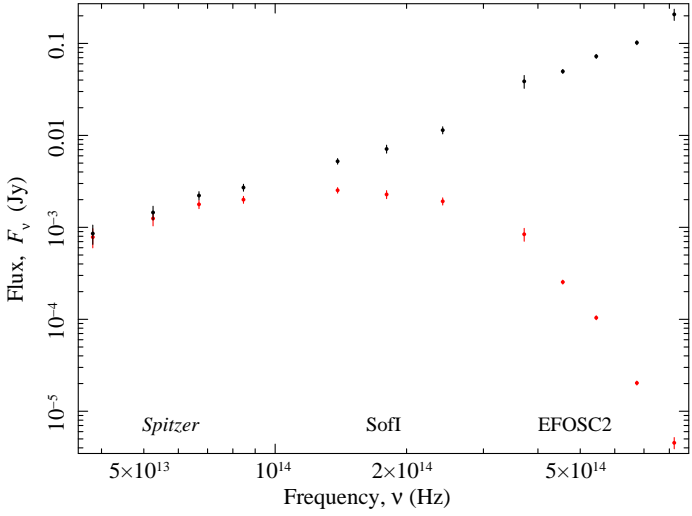}}
  \caption{Flux versus frequency plot for \xtesixteen: uncorrected (lower) and corrected (upper) for Galactic extinction in that direction of  $E_{B-V} = 2.143$.} 
  \label{fig.xte1637_SED} 
\end{figure}

Figure\,\ref{fig.xte1637_SED} shows the spectral shape of the counterpart of \xtesixteen, including the \glimpse\ data. Also plotted are the fluxes corrected for a Galactic extinction in that direction \citep{schlegel1998:ApJ500} of  $E_{B-V} = 2.143$\footnote{Note that these extinctions should be treated with caution as estimates so close to the Galactic plane ($<5\deg$) are unreliable.\label{noteX}} ($A_K \sim 0.8$, $A_V \sim 7.1$; \citealt{cardelli1989:ApJ345}), though this should be considered as an upper limit. 
Though we are unable to find a single model -- for any value of Galactic extinction -- that fits both the NTT and \glimpse\ data, we find that the NTT data alone can be fit by black body radiation at a temperature of 3100--12800\,K and corresponding extinctions of  $0.4 < E_{B-V} < 2.0$. However, when extended to the \spitzer\ data, this model underestimates the flux at those wavelengths. In fact, the \spitzer\ data itself implies a power law ($F_{\nu} \propto \nu^{\alpha}$) of spectral index $1.0 < \alpha < 1.7$ for the same range of extinctions as implied by the black body fit to the NTT data. A power law of similar slope is not a good fit to the NTT data alone or the NTT and \spitzer\ data combined ($\chi_{\nu}^{2} >> 1$).  
This implies that there is excess emission at the \spitzer-IRAC wavelengths (see section \ref{section:conclusions}), either due to intrinsic IR excess  or an excess at the time of the \spitzer\ observations which, as catalog magnitudes, were not simultaneous with the NTT observations.

\subsection{\igrseventeen}

Outbursts from \igrseventeen\footnote{While \cite{Chelovekov2006:AstL.32} and \cite{Chelovekov2010:AstL.36} refer to the source by differing IGR designations (IGR\,J17364$-$2711 and IGR\,J17380$-$3749), herein we shall adhere to that of the \integral\ catalog \citep{Bird2007:ApJS170,Bird2010:ApJS186}.}
were first reported from the \emph{International Gamma-Ray Astrophysics Satellite} (\integral) by \citet{Chelovekov2006:AstL.32} with a fuller investigation of \integral\ and other observations presented in \cite{Chelovekov2010:AstL.36}. 
\cite{Chelovekov2006:AstL.32} classified the source as an X-ray burster -- a weakly magnetized, accreting neutron star in a \lmxb\ system -- on the basis of a type-I burst or a thermonuclear explosion, due to the unstable burning of H and He on the neutron star surface \citep{Hoffman1978:Natur.271}. 
During the 2008 outburst the X-ray position was refined to arcsecond precision (Table\,\ref{table:positions}; \citealt{Krimm2008:ATel.1714}) by the \swift-XRT. 

\begin{figure} 
  \centering 
  \resizebox{\hsize}{!}{\includegraphics[angle=-90]{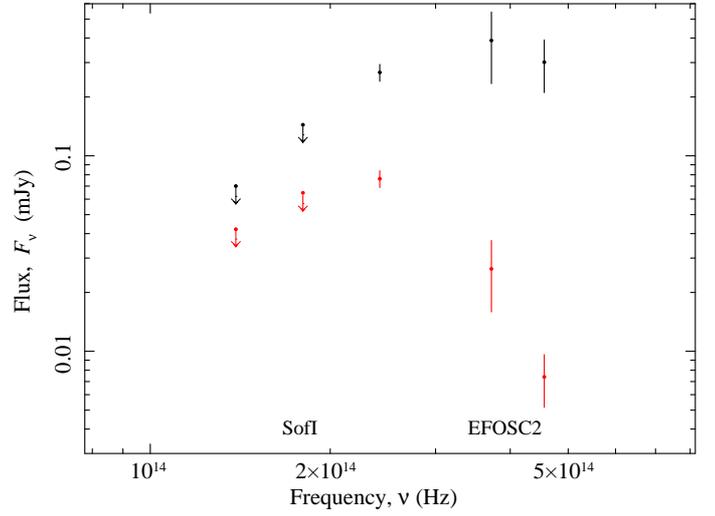}}
  \caption{Flux versus frequency plot for \igrseventeen: uncorrected (lower) and corrected (upper) for Galactic extinction  in that direction of  $E_{B-V} = 1.506$.} 
  \label{fig.igr17379_SED} 
\end{figure}

At the X-ray position a new source is detected in our $i$ and $R$ images on both nights (Figure\,\ref{fig.4fields};  best seeing was $\sim 1.4$\arcsec\ in nIR filters and $\sim 1.2$\arcsec\ in optical filters). The source exhibits no variability ($<0.05$, $<0.1$ magnitudes respectively) between the 2 epochs which are separated by 3 days (we present the average magnitudes). The lack of variability is not surprising since observations were taken after the 2008 September 2--21 X-ray outburst had faded \citep{Markwardt2008:ATel.1709}. The source is also detected in the second epoch $J$ band image but not in the first, though the magnitude limit on that night is consistent with the detected magnitude. We note that the counterpart is at the edge of the psf of a nearby, bright 2MASS object (J17375858$-$3746228; $J = 13.210$, $H = 12.067$, $K_s = 11.743$).  
There was no detections in the  $U,$ $B$ and $V$ bands and  the lack of suitable optical standard stars on either night have lead us to omit their magnitude limits. 
The position of the optical/nIR counterpart was derived as RA, Dec = 17:37:58.77 $-$37:46:19.9  ($\pm 0.32$\arcsec) from the $R$ and $J$ band images. 
The spectral shape of the proposed OIR counterpart of \igrseventeen\ (Figure\,\ref{fig.igr17379_SED}) is not particularly constraining but is consistent with stellar-like black body emission for a realistic range of temperatures, at extinctions up to the value  for Galactic extinction in that direction \citep{schlegel1998:ApJ500} of  $E_{B-V} = 1.506^{\ref{noteX}}$ ($A_K \sim 0.5$; \citealt{cardelli1989:ApJ345}). This region is not covered by the \glimpse\ survey so we are unable to extend the SED redwards.

\subsection{\igrseventeenfive}

 \begin{figure} 
  \centering 
  \resizebox{\hsize}{!}{\includegraphics[angle=-90]{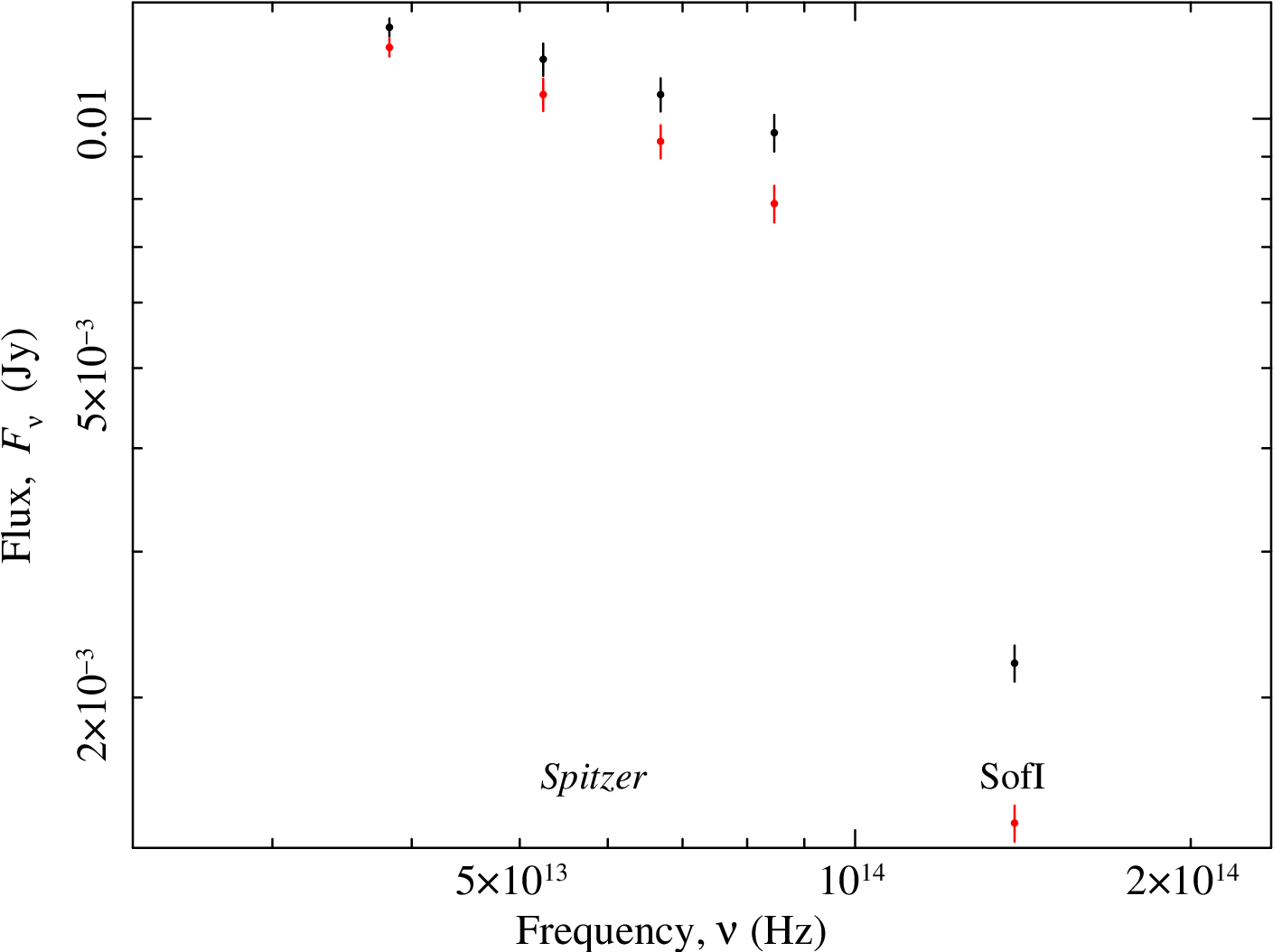}}
  \caption{Flux versus frequency plot for \igrseventeenfive: uncorrected (lower) and corrected (upper) for Galactic extinction in that direction of  $E_{B-V} = 1.302$.} 
  \label{fig.igr17585_SED} 
\end{figure}

Detected by \integral\ \citep{Bird2007:ApJS170,Bird2010:ApJS186}, \igrseventeenfive\ is assumed to be a \lmxb\ based on Galactic position and the lack of an alternative classification \citep{Revnivtsev2008:A&A.491}. Monitored as part of  both the \integral\ all-sky survey  in hard X-rays \citep{Krivonos2007:A&A.475} and the \swift-BAT 58-Month Hard X-ray Survey \citep{Baumgartner2010HEAD} there has been no evidence of flux variation. No further analysis of this source was published so its true classification remains unknown, though an accurate X-ray position was obtained by \chandra\  (Table\,\ref{table:positions}; M. Revnivtsev, private communication).

At the X-ray position we find a bright, $K_S = 14.183 \pm 0.05$ counterpart  in our single SofI image of the field (Figure\,\ref{fig.4fields}; seeing $\sim 0.8$\arcsec) and we derive a position of RA, Dec = 17:58:29.83 $-$30:57:01.6  ($\pm 0.1$\arcsec). This coincides with a \glimpse\ catalog source (G359.6862-03.4261), the magnitudes of which we reproduce in Table\,\ref{table:mags}. 
From the SED (Figure\,\ref{fig.igr17585_SED}), it is clear that the extrapolation of the \spitzer\ data overestimates the  $K_S$ magnitude significantly. In fact the \spitzer\ data implies a spectral index of $\alpha = -0.37 \pm 0.10$ (when  optical extinction is set to that of  Galactic extinction in that direction \citep{schlegel1998:ApJ500}, $E_{B-V} = 1.302^{\ref{noteX}}$; $A_K \sim 0.4$; \citealt{cardelli1989:ApJ345}) but then drops by almost an order of magnitude, requiring a steepening to $\alpha \gtrsim 5$ to agree with the $K_S$ flux. 
This implies either intrinsic excess emission at the \spitzer-IRAC wavelengths or it is possible that the source is variable, though without multiple photometric observations, we cannot test this.

\subsection{\gxnine}

The persistent X-ray source \gxnine,  a.k.a. \sgrxthree, was initially discovered by sounding rockets \citep{Bradt1968:ApJ.152,Mayer1970:ApJ.159} and later confirmed by the first dedicated X-ray astronomy satellite, Uhuru (as 2U\,1758-20; \citealt{Giacconi1971:ApJ.165,Giacconi1972:ApJ.178}).  
Thereafter, it was observed by most other X-ray missions which revealed that the source is likely an `Atoll' source, a class of neutron star \lmxb\ (e.g., \citealt{Langmeier1985:SSRv.40,Hasinger1989:A&A.225,Schulz1989:A&A.225}). Previous searches for an optical or nIR counterpart \citep{Gottwald1991:A&AS.89,Naylor1991:MNRAS.252} did not yield any plausible candidates due to the lack of an accurate X-ray position, which is now available from \chandra\  observations (Table\,\ref{table:positions}; M. Revnivtsev, private communication).

At the X-ray position we detect a $K_S = 15.35 \pm  0.04$ source  in our single SofI image of the field (Figure\,\ref{fig.4fields}; seeing $\sim 1.4$\arcsec) and we derive a position of RA, Dec = 18:01:32.15 $-$20:31:46.2  ($\pm 0.16$\arcsec). We note that there is also a source directly to the West which appears in the \glimpse\ (G009.0755+01.1545) and 2MASS (J18013196$-$2031464; $K_S = 13.263 \pm  0.055$) catalogs but we find that it is $>6\sigma$ from the X-ray position and is hence likely unrelated to \gxnine.


\section{Discussion and conclusions}\label{section:conclusions}

To approximate the probability of a chance superposition of the 90\% X-ray positions with random sources in the respective fields, we calculate $P \approx \rho_N \times A_{Err}$; where $\rho_N$ is the surface area number density of observed sources down to the limiting magnitude and $A_{Err}$ is the area of the X-ray positional error. 
The probability of chance coincidence for \xtesixteen\ and \igrseventeen\ are  quite high at 86\% and 28\% respectively, due to a large number of sources in the deep fields but more significantly, the radius of the X-ray positional errors. 
The importance of sub-arcsecond positions from e.g. \chandra\ is further illustrated by the significantly lower probabilities of \igrseventeenfive\ and \gxnine\ (8\% and 7\%) which have comparable number densities to the previous two fields but much more accurate X-ray positions. The probability of chance superposition could be reduced further if we limited our search to sources of specific magnitude ranges but due to a lack of information on these sources, we make no such assumption.

On the basis of positional coincidence with the arcsecond accurate X-ray positions, we suggest unambiguous (in so far as there is only one per source) counterpart candidates for four \lmxbs\ near the Galactic centre. We caution that given the probability of chance superpositions in these crowded fields, the associations are tentative and require further investigations of e.g., spectral type, variability, distance, to confirm the associations with the X-ray sources. 
The spectral shapes of the counterpart candidates are not particularly constraining but, in general, could be consistent with stellar-like black body emission in the optical/nIR but with an excess at the longer, \spitzer, wavelengths, as for \xtesixteen. The excess could be due to warm circumbinary dust, likely related to the formation of the compact object (e.g., \citealt{Rahoui2010:ApJ.715}) or to non-thermal, synchrotron jet emission (e.g., \citealt{Gelino2010:ApJ.718}).
Further, possibly simultaneous, photometric and spectroscopic observations are required to disentangle the different emission components and identify the spectral type of the companion star, if it is  detected over the other possible emission mechanisms.


\begin{acknowledgements}  
We thank the referee for their constructive comments and  M. Revnivtsev for supplying \chandra\ positions for two of the sources in our study. 
This work was supported by the Centre National d'Etudes Spatiales (CNES) and is based on observations obtained with MINE: the Multi-wavelength INTEGRAL NEtwork.
This research has made use of NASA's Astrophysics Data System and the VizieR catalogue access tool, CDS, Strasbourg.
%
IRAF is distributed by the National Optical Astronomy Observatory, which is operated by the Association of Universities for Research in Astronomy (AURA) under cooperative agreement with the National Science Foundation.
GAIA was created by the now closed Starlink UK project funded by the Particle Physics and Astronomy Research Council (PPARC), and has been more recently supported by the Joint Astronomy Centre Hawaii funded by PPARC and now its successor organisation, the Science and Technology Facilities Council (STFC). 
\end{acknowledgements}

\end{document}